\documentclass[12pt]{article}
\usepackage{mathtext}
\usepackage{graphicx}
\usepackage[utf8]{inputenc}
\usepackage{amsfonts}
\usepackage{amssymb}
\usepackage{amsmath}
\numberwithin{equation}{section}

\newcommand{\la}{\lambda}

\begin{document}
\title{About quantum computer software}

%\title{Квантовый компьютер: неопределенность ''сложность - точность''}

\author{Y.I.Ozhigov\thanks{ozhigov@cs.msu.su}\\
Moscow State University \\
of M.V.Lomonosov, VMK Faculty,\\
Institute of Physics and Technology RAS}
\maketitle
%\tableofcontents

\begin{abstract}
Quantum computer is the key to controlling complex processes. If its hardware, in general is successfully created on the basis of the physical baggage of the 20th century, the mathematical software is fundamentally lagging behind. Feynman's user interface in the form of quantum gate arrays, cannot be used for the control because it gives the solution of the Schrödinger equation with quadratic slowdown compared to the real process. The software must then imitate the real process using appropriate program primitives written as the programs for classical supercomputer. The decoherence will be reflected by some constant - the number of basic states that can fit into the limited of memory available to software. The real value of this constant can be found in the experimental realization of Grover search algorithm. Rough estimates of this constant are given based on the  simplest processes of quantum electrodynamics and nuclear decay.  
\end{abstract}

\section{Introduction and background}

Over the past 20 years, the quantum computer project has evolved from a ''storm and onslaught'' into a long-term project that engages an increasing number of researchers in various fields. As the experimental work revealed the true value of the decoherence problem, which in the late 90's was considered as technical, there was a growing need for a deeper understanding of what we mean by ''quantum computer'', and how it should be created. In particular, we believe that the project of its construction should be based not only on the achievements of the traditional quantum physics, which mainly dealt with relatively simple systems and processes, but also on the ideology of computing and real computers, especially since this ideology in the field of complex processes has given us greater opportunities than the analytical technique of quantum theory of the past.

The experience gained in modern physics, makes it possible to be optimistic about the possibility of creating an experimental sample of hardware for a quantum computer in the near future. However, the computer does not have only hardware, it needs the software - the operating system. If for classical computing, the creation of an operating system is the same large-scale task as the construction of the physical part, the quantum operating system is much more difficult, since hardware itself depends on it.

We can schematically represent a quantum computer in the form of a three-level structure, where the lower level will be its quantum processor, whatever it is, the second level will be the system part of the software - the drivers of quantum devices and the general program for controlling the computation, and the upper level will be the user interface directly interacting with a human (see figure 1).

\begin{figure}
\begin{center}
\includegraphics[scale=0.7]{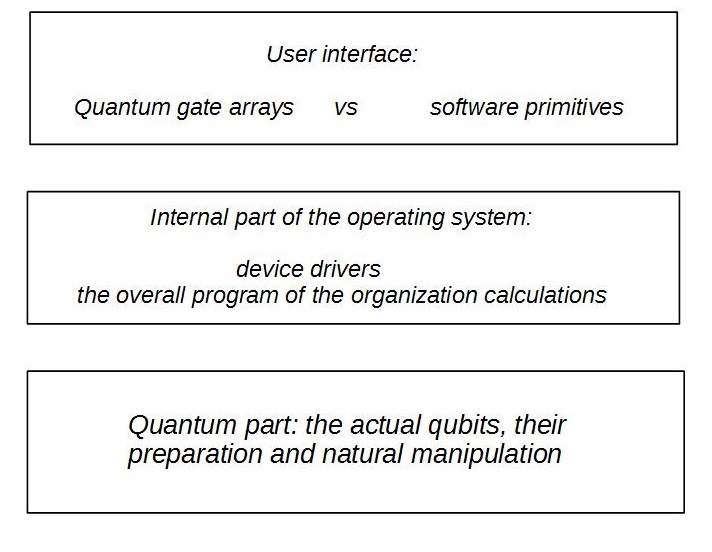}
 \caption{Scheme of quantum computer}
\end{center}
\end{figure}

R. Feynman in \cite{Fe3} proposed a user interface based on an array of quantum gates implementing the simplest unitary operators on a small number of qubits. This interface is aimed at the detailed computation of the wave function, the scheme of which was developed in \cite{Za}. This quantum scheme differs from the direct solution of the Shrödinger equation on a classical computer only in that the amplitudes $\la_j$ of the current quantum state $|\Psi (t)\rangle$ are not calculated directly, but are modeled by the unitary dynamics of quantum state $|\Psi(t)\rangle=\sum\limits_{j=0}^{N-1}\la_j|j\rangle$ in $n$ qubit space spanned by classical basic states $|j\rangle=|0\rangle,|1\rangle,...,|N-1\rangle$, $N=2^n$. 

The operator of unitary evolution can be represented by Trotter approximate formula
\begin{equation}
exp(-\frac{i}{\hbar}(E_{kin}+V)t)\approx (exp(-\frac{i}{\hbar}E_{kin}dt)exp(-\frac{i}{\hbar}Vdt))^{t/dt}
\label{trot}
\end{equation}
where the error of the single expansion of exponents with degree $dt$ is of the order $dt^2$ (it would suffice to expand exponentials up to the first degree of $dt$). We have to realize separately these two operators, corresponding to kinetic and to potential energy. 

It is assumed that the real one-dimensional space of classical states is at first translated by a linear transformation into the segment $[0,\sqrt{N}]$, which is then discretized by the qubit representation of numbers with an approximation accuracy of $1/N$: $x_k\approx k/ \sqrt{N},\ k=0,1,...,N-1$. Such a representation of the wave vector requires an appropriate discretization of the operators. The coordinate operator then becomes a diagonal matrix with the numbers $V (x_0),V(x_1), V(x_2),...,V(x_{N-1})$ on the main diagonal, the momentum operator is represented as $FT^{-1}\ P\ FT$, so that where the representation of the kinetic energy operator (in the space of its eigenvectors) is also diagonal, with the numbers \newline $exp(p_0^2/2m), exp( p_1^2/2m), exp( p_2^2/2m),..., exp (  p_{N-1}^2/2m)$ on the main diagonal, and the Fourier transform $FT$ is represented as a $n$ - qubit quantum Fourier transform of the form
\begin{equation}
\label{QFT}
\begin{array}{ll}
&QFT:\ |a\rangle\rightarrow \frac{1}{\sqrt{N}}\sum\limits_{c=0}^{N-1}exp(-2\pi i ac/N)|c\rangle,\\
&QFT^{-1}:\ |c\rangle\rightarrow \frac{1}{\sqrt{N}}\sum\limits_{a=0}^{N-1}exp(-2\pi i ac/N)|a\rangle
\end{array}
\end{equation}
Then the part of evolution corresponding to the exponential of the potential energy operator $exp (- i/\hbar Vt)$ for the simple form of the potential will be realized as a quantum subroutine, the quantum Fourier transform can also be realized by Shore scheme (\cite{Sh}), and the operator corresponding to the kinetic energy and time $t$ can also be realized as a quantum subroutine.
We can transfer the coordinate basis to impulse basis using $QFT$; for example, if the real impulse $p_k\in[-\sqrt{N}/2,\sqrt{N}/2]$ this transformation has the form $QFT\ diag(e^{\pi i\ a})_{a=0,1,...,N-1}$.

  Applying the Trotter approximation, we obtain a model of unitary dynamics with quadratic slowdown compared to the real process (see \cite{Za}). Really, the complexity of method (\ref{trot}) as the number of steps in the modeling is about $t/dt$, if we fix the admissible error of the resulted state by $\epsilon=dt^2t/dt=tdt$, the resulting complexity will be  $t^2/\epsilon$.

This method of detailed modeling requires memory growing linearly with the number of real particles, but cannot be used to control a complex system, because it involves a-priory modeling of the process with the transfer of the result to a new similar process, whereas in reality any complex process is not exactly reproducible, and therefore the control requires modeling in real time.

So, the Feynman interface has natural limits of applicability which significantly narrow our opportunities in management of difficult processes of the quantum nature.

The second basic lack of the Feynman interface: it is aimed to only modeling of unitary dynamics of pure states, whereas the case of real systems requires additional decoherence suppression. Quantum error correction codes (\cite {ShC}) only work if a strict condition is met: a quantum processor must work well without them on several hundred qubits, which is problematic in itself. On the other hand, the decoherence as the influence of environment in itself rests on  methodologically dark place in quantum theory - a problem of wave function collapse. The quantum master equation, or Kraus operators (\cite {BP}) applied to the detailed description of decoherence bring us besides to the same shortcoming again. Besides, the states important in the detailed modeling are especially sensitive to decoherence; it makes such a way problematic.

These shortcomings of traditional approach induce us to search the more practical scheme of modeling of reality on the quantum computer which would combine theoretical representations of quantum physics with time of real processes. The quantum operating system intended for control of real process should work in real time that assumes usage of special program primitives instead of gate arrays: the subroutines imitating real process on the quantum computer. These primitives depend on the considered process; here we will touch only chemical primitives which should be used when modeling the processes belonging to electrodynamics: associations- dissociation of atoms and their interaction with a field. In other areas, perhaps, other primitives will be required.

In this case, the operating system control commands must themselves reproduce the corresponding physical process in some limited form, concerning a few real particles. An operating system, like a program written for a classical supercomputer, must produce a process that has a high degree of adhesion to the simulated real process. Only in this case we can expect to receive a quantum computer as a working tool.

This requirement of similarity of control and simulation is absent in classical computers because classical physics allows the same description of the observer and the observed system, so that observation does not change its state. It does not take place in quantum case, and therefore some degree of reproducibility of the quantum model in the classical operating system is necessary. We do not know exactly how quantum physics works in complex processes, and therefore it is necessary to ensure our model with classical means of displaying reality. 

\section{Purpose and capabilities of a chemical quantum computer}

A chemical quantum computer must simulate chemical reactions in real time in order to control such reactions. It plays the same role in relation to reactions as x-ray diffraction analysis plays in relation to the stationary structure of molecules, namely, the quantum computer must allow to see the mechanisms of reactions taking into account the electromagnetic field, which will mean full control over chemistry. This problem is real only for relatively simple reactions that take a short time. For complex chemistry, this task can be realized only in the context of a living organism, where the gigantic uncertainty of the initial conditions of chemistry is radically harshly reduced by the structure of proteins, that is, ultimately, by DNA.

We show the principal possibility of tracking a chemical reaction using a classical transistor included in the modern chip.  The number of operations does not exceed the number of $n$ passes of light through the transistor: the diameter of one transistor is not less than $10\ nm\approx 10^{-6}\ cm$. How fast can it work? 
The number of operations does not exceed the number $n$ of passes of light through the transistor: $c\approx 3\cdot 10^{10} \ cm/sec,\ n\approx 3\cdot 10^{16}$.
Time step $dt$ is needed for confident dynamics modeling. Its value is found from the uncertainty relation $dE\ dt = \hbar\approx 10^{-27}\ erg\ sec$.
For electrodynamics $dE\approx 10^{-17}\ erg,\ dt\approx 10^{-10}\ sec$, and one transistor copes with modeling of evolution of one charge; even if to consider transition of an electron from atom to atom which time is approximately $10^{-12}\ sec$, one transistor will, in principle, manage to reproduce a state of such electron.

For nuclear physics, $dE\approx 10^{-5}\ erg,\ dt\approx 10^{-22}\ sec$, and a single transistor cannot even come close to the required operating speed to simulate the dynamics of a single nucleus, even with the classical representation of its dynamics. In addition, predictive modeling requires quantum mechanics, where one atom $^{235}U$ requires a memory of at least $2^{235}$.

Thus, a quantum computer should, in principle, be able to handle simulations of simple chemistry, at least in terms of clock speed, but not nuclear processes at the quantum level. Nuclear transformations are beyond the reach of computer technology based on electromagnetism; we will come to the same conclusion in a different way below.

\section{Software primitives for chemical computer}

Quantum parallelism in controlling the dynamics of chemical reactions can be realized through a quantum computer on the charge states of electrons in solid-state quantum dots or in a system of atoms placed in optical cavities. This scheme has been investigated in a number of papers (see, for example, \cite{Tsu}), here we describe briefly the software primitives needed to implement this chemical type of quantum computer. 

The states of the real set of atoms and the electromagnetic field are represented in this scheme as follows: atoms are represented by quantum dots - artificial ''atoms'' connected to each other by optical fibers and conductors, so that both photons and electrons can move from dot to dot. Such a system should contain the following software primitives corresponding to elementary scenarios: the transition of an electron from level to level at a given dot with simultaneous absorption or emission of a photon, the transition of a photon from one point to another, the transition of an electron from one point to another, as well as the exchange of photons and electrons between dots along an optical fiber. Some arrays of standard quantum gates correspond to such scenarios, for example, the movement of a photon from point to point corresponds to the SWAP operator applied to two qubits (see figure 2).

\begin{figure}
\begin{center}
\includegraphics[scale=0.7]{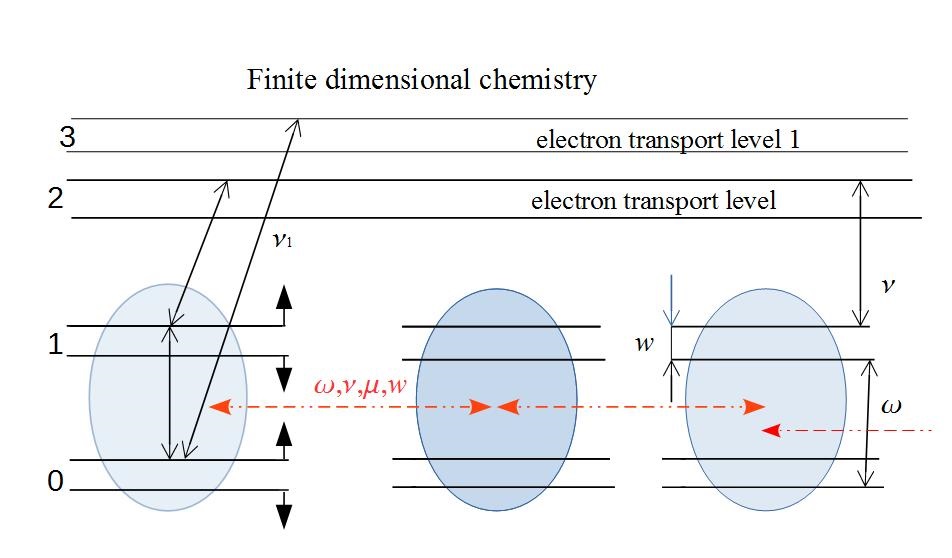}
 \caption{Chemical quantum computer}
\end{center}
\end{figure}

The Hamiltonian corresponding to this dynamic can be written out in the same way as for the finite-dimensional Tavis-Cummings QED model (see \cite{JC},\cite{TC}). This model can be used to represent the reactions of association - dissociation of atoms, because there is a spin dynamics, and the movement of electrons from an atom to another atom depends on the spin state of the electrons available at these points - according to the Pauli principle. This model reflects the features of quantum charge and field dynamics, which no classical computer can reflect: entanglement and nonlocality.

Hardware for such a model can be made in the near future on solid-state quantum dots. A more accurate version of a quantum computer can be obtained using optical resonators with multilevel atoms placed inside - this is a multilevel modification of the Tavis-Cummings-Hubbard model (see \cite{TC},\cite{Hub}). The variant with optical cavities, where atoms are in a vacuum and are not exposed to heat, is actually described by the first principles of quantum theory, whereas the solid-state model requires approximate methods, here quantum dots are surrounded by a huge number of other atoms. Therefore, cavities are preferable to accurately display all decoherence factors of the cavity. 

In the described chemical model of a quantum computer, operations corresponding to standard quantum gates, such as CNOT, Toffoli and single-qubit gates, can be performed approximately. However, the realization of gates is the demonstration, whereas chemical primitives are the working apparatus needed to simulate real complex chemical reactions.

\section{General program controlling quantum computation}

The core of the quantum operating system is a program written for a classical supercomputer, which should control the general course of computation. If software primitives with high accuracy simulate the dynamics of a small fragment of the entire model, the operating system kernel must control the dynamics of the quantum state of the entire system under consideration, and for a sufficiently large period of time, in order for the model to reflect the meaningful features of the real system.

Here we come to the point where our knowledge of the microcosm diminishes. We don't know how quantum theory works in complex processes. We have only a mathematical scheme of quantum algorithms - in the Feynman interface and the assumption of the existence of actual infinities, such as the exponential dimension of the Hilbert space of quantum states.

Quantum computer thus becomes an experimental device that tests the boundaries of mathematical abstractions of analysis, when we know for sure that these abstractions only approximately correspond to reality - for example, the impossibility to normalize the eigenfunctions $c\cdot exp(ipx/\hbar)$ of the momentum operator $-i\hbar\nabla$. The mathematical correctness of quantum mechanics is achieved, as is known, only in the discrete representation of the space of classical states of the particle as points of the form $k / N,$ where $k=0,1,...,N-1,\ N=2^n,$ for $k$, which binary decomposition represents the basic vector $|k\rangle$ in $n$ qubit space. Then, by linear transformation in the classical space, we translate the segment $[- A, A]$, on which the dynamic scenario is played out, into the set of points $k/N$, and the transition to the momentum basis is made in the form of a quantum Fourier transform.  In this case, we have $|\Psi\rangle= \sum\limits_{j=0}^{N-1}\la_j|j\rangle$ and can apply the Feynman gate interface to simulate the wave vector dynamics according to the above scheme.

However, as we have seen, this is not enough for control at the quantum level. We must take the next step, further narrowing the standard quantum formalism. The operating system must be able to simulate quantum evolution over a period of time, at least approximately, without the participation of the quantum part of the computer. There should not be the gap between classical computation and the quantum process that is assumed in the Feynman scheme of the quantum computer, because the experiments clearly speak in favor of the impossibility of direct scaling of this scheme. We can expect to succeed only if we observe a smooth transition from quantum dynamics modeling on a classical computer to quantum modeling.

For the classical simulation of quantum dynamics, there is one transparent limitation, which we will consider below.

\section{Quantization of amplitude}

In the representation of amplitudes in classical computation, they are always quantized, that is, they have the form:
\begin{equation}
\label{lambda}
\la_j=(k_j+il_j)\epsilon,
\end{equation}
where $\epsilon$ is the small nonzero value - amplitude quantum, $k_j,\ l_j$ are natural numbers. This representation of amplitudes follows from the linearity of quantum theory. It also requires an appropriate choice of classical basic states, but due to the smallness of $\epsilon$ this does not lead to any revision of the experimentally confirmed part of the quantum theory, and will only affect the scaling of the quantum computer.

So, the quantum operating system considers only states of the form
\begin{equation}
\label{state1}
|\Psi\rangle=\sum\limits_{j\in J}\la_j|j\rangle,
\end{equation}
where the amplitudes have the form \eqref{lambda}. So, the summation in \eqref{state1} extends to at most $1/\epsilon^2$ summands of the set $J=\{ 0,1,...,N-1\},\ N=2^n$ of basis states of the $n$ qubit system. The maximum value of $n$ is built into the operating system, so we must consider this number as the physical constant regardless of what happens in the real system. Similarly, the minimum value of the amplitude resolution constant $\epsilon$ is also built into the operating system, and we must consider this constant as physical. 

This makes adjustments to the traditional interpretation of the quantum computer and gives new meaning to the experiments on its creation. Note also that an operating system without a quantum part cannot reproduce real evolution. The factor of instantaneous action at a distance - nonlocality, which leads to the violation of Bell inequality, has a fundamentally quantum nature and cannot be reproduced in an operating system based on classical physics (on the use of nonlocality, see, for example, \cite{Oz1}).

\section{On the size of the amplitude quantum}

We do not know what is the value of $\epsilon$ in reality, so we give only rough estimates of this number, based on the experience of quantum description of real processes.

Let $M$ be the set of qubits of the system, consisting of $n$ qubits, and $|\Psi\rangle$ be any quantum state of these qubits. It is called non-entangled if there is such a partition of $M=M_1\cup M_2$ into two disjoint non-empty sets and states $|\Psi_1\rangle,\ |\Psi_2\rangle$ on these sets such that $|\Psi\rangle=|\Psi_1\rangle\otimes |\Psi_2\rangle$. Otherwise, the state $|\Psi\rangle$ is called entangled. The complexity of the state $|\Psi\rangle$ on the set $M$ is the qubit size of its maximum entangled tensor divisor, that is, the maximum of the natural numbers $s$ such that there is a subset of $M_1\subseteq M$ and the states $|\Psi_1\rangle,\ |\Psi_2\rangle$ on $M_1$ and $M-M_1$ respectively, such that $|\Psi\rangle=|\Psi_1\rangle\otimes |\Psi_2\rangle$, $m_1$ contains $s$ elements and $|\Psi_1\rangle$ is entangled. This state $|\Psi_1\rangle$ is called the quantum kernel of the state $|\Psi\rangle$, and the corresponding set $M_1$ is the kernel carrier.

There can be several kernels, since the maximum number of $s$ from the definition can correspond to different sets of $M_1$ qubits. Naturally, this definition may depend on very small amplitudes, so that the complex state may be very close to the simple one. However, if we consider only states whose amplitudes $\la_j$ have the form \eqref{lambda}, this proximity will be limited by the value $\epsilon$. From the following it will be clear that it is impossible to aim $\epsilon$ to zero for complex systems, and therefore the complexity is determined in this way correctly. We will denote the complexity of the state $|\Psi\rangle$ by $C (\Psi)$.

Let $\tau\in S_N$ be a permutation of the basis vectors of state space corresponding to the set of qubits $M$. Then the state $\tau |\Psi\rangle$ is called a quasi-partial representation of the state $ | \Psi\rangle$. For example, for a set of $ n $ harmonic oscillators, their typical basic state is $(q_1, q_2,..., q_n)$, its Fourier transform of the form $Q_k=\alpha\sum\limits_jq_je^{- \beta \ ikj}$ means a transition to the phonons - quasiparticles with new coordinates $Q_k$.

Another example: a generalized GZH state of the form $\frac{1}{\sqrt 2} (|00...0\rangle+|11...1\rangle)$ in which all $n$ qubits are entangled, but it can be reduced to the non entangled  state by successive operations $CNOT$, which are permutations of the basis vectors of the space.

The absolute complexity $a (\Psi)$ of the state $|\Psi\rangle$ is the minimal complexity of all its quasi-partial representations. Formally:

\begin{equation}
\label{abs_complexity}
A(\Psi)=min_{\tau\in S_n}C(\tau |\Psi\rangle).
\end{equation}
 
Absolute complexity is the number of qubits required to represent the quantum kernel of a given state. The state space in which this kernel lives thus has the dimension $2^{A (\Psi)}$, which we will further denote by $N$.

We assume that for the state \eqref{state1} the amplitudes of its components always have the form \eqref{lambda} for some $\epsilon>0$. 
It is the same as the amplitude will take only four values, $\pm\epsilon,\ \pm i\epsilon$, and the states of the set $J$ in the expansion \eqref{state1} take some distribution in the classical space ${\cal K}$ configurations of the system, so if we want to calculate the ''wave function'' in the usual sense of the term, we must sum the amplitudes for small sections of $\delta_k{\cal K}$ of the space ${\cal K}$ so four types of ''amplitude quanta'', adding in all the states of the $J\cap \delta_k{\cal K}$, give the value of the ''wave function'' in the center of the chunk $\delta_k{\cal K}$.

The length of the recording of the spatial position and amplitude of any basic state varies as a logarithm of the total memory required to store the entire wave function in the operating system. Therefore, we can approximately assume that the total memory $ Q$ of the operating system coincides with the maximum dimension of the Hilbert state space of the simulated ensemble. Then we get $\epsilon = 1/\sqrt{Q}$. The value $\epsilon$ must thus be considered a dimensionless constant, since the dimension of the physical quantity will refer to the basis states $|j\rangle$ in \eqref{state1}.

The representation of the wave function with $N$ basic states is achieved in the only case - when all amplitudes are modulo $\epsilon$. If the basic states are less than $Q$, it means that the quantum amplitude is summed on chunks $\delta_k{\cal K}$, and then we have the uncertainty relation ''complexity-accuracy'' of the form
\begin{equation}
N\ log_2(1/\varepsilon)\leq Q
\label{uncertainty}
\end{equation}
where $\varepsilon$ is the accuracy of representing the amplitudes for a traditional ''wave function'' for a set of $N$ points in space ${\cal K}$. So, for values $N\ll Q$, the accuracy of the representation of the traditional ''wave function'' will be so high that it will be impossible to distinguish it from the analytically obtained expression; for simple systems this is the case.

Let's make one remark about contact with the environment. If it has no long-term memory, it is described as an open quantum system using the quantum master equation. Such a Markov quantum process is already described by the density matrix, which we can represent in the form $\rho=\sum\limits_jp_j|\phi_j\rangle\langle\phi_j|$ for orthonormal system of $|\phi_j\rangle$ .  Since there is no coherence between these terms, their modeling can be classically distributed over different processors, which does not concern the constant $Q$. Therefore, we discuss here only the unitary dynamics of the pure state.

\section{Experimental determination of constant $Q$}

The dimensionless constant $Q$ is the maximal number of basic states in which quantum kernel of any system lives. It thus has a physical nature, and is subject to experimental search. To do this, it is necessary to distribute the amplitude over a very large number of classical states of some ensemble, so that the presence of the amplitude quantum $\epsilon$ would lead to a gross deviation from coherent dynamics, which could be recorded in the experiment. If we deal with well-studied physical ensembles, it is very difficult to do so, since they are amenable to study by standard means precisely because their absolute complexity is small and therefore the accuracy of determining the amplitude can be very high.

The most reliable way is to implement the Grover search algorithm - GSA  (see \cite{Gr}) for as many $n$ qubits as possible. We have to reproduce the computation from this algorithm to find the root of the equation

\begin{equation}
\label{grov}
f(x)=1, 
\end{equation}
where $f$ is a Boolean function of $n$ variables such that this equation has a unique solution $x_t$.

Note that after $s$ applications of Grover's $G$ operator, the state will have the form $cos(2s/\sqrt{N})|\tilde 0\rangle+sin(2s/\sqrt{N})|x_t\rangle$ with high accuracy, where $|\tilde 0\rangle=\frac{1}{\sqrt N}\sum\limits_{j=0}^{N-1}|j\rangle$. This specific form, when the amplitude of a single state exceeds the amplitude of all other states, in which it is the same and nonzero, will remain at any permutation of the basic vectors $\tau$, so starting from the first step all $n$ qubits will form a single quantum kernel, and the maximum number of $n$ qubits, for which GSA will give the correct answer, will give the value $Q=2^n$. Here we assume that GSA works if at least for the first application $G$ the amplitude of the target state will exceed all others; to fix this, it is necessary, of course, to make numerous measurements of the result of one such application, which with a large value of $Q$ can become a problem. Anyway, the experiment can estimate $Q$ from below if you try to lead the GSA to the end point by making $[\pi\sqrt{N}/4]$ applications of $G$.

\section{Physical sense of constant $Q$}

We consider two processes: the state transition of an electron in an atom $Rb^{85}$ and the decay of an unstable nucleus $He^6$. The first process is described by quantum electrodynamics quite accurately, a complete quantum description of the second is not yet available.

We will proceed from the criterion of accurate drawing of the wave function, when each step of its computer description requires one new basic state. This follows from the speed of the quantum walk, in which the wave front propagates at a linear speed (as opposed to the classical walk, in which the speed is proportional to the square root of time). Let $t$ be the total time of the process, $dt$ is the step of the computer description of this process in time, then the number of basic states required for ''accurate drawing'' of the process is $N=t/dt$. The values of $t$ are determined experimentally, and $dt$ is derived from the energy - time uncertainty relation.

For Rabi oscillation of a rubidium atom occurring with the emission of a photon with a wavelength of approximately $1.4\ cm$ we have: $ \omega \approx 10^{10}\ sec^{-1}$, $E_{QED}=\hbar\omega\approx 10^{-17}$, $dt\approx\hbar/E_{QED}=10^{-10}$. Given the time of Rabi oscillation $t\approx 10^{-6}\ sec$, we get $N=t/dt\approx 10^{4}$. Thus, if this process can be ''well drawn'' in quantum theory, then $Q\geq 10^{4}\approx 2^{13}$ and GSA must work on about 13 qubits, which seems quite real.

Now we consider the decay of the nucleus of Helium-6 isotope: $He^6\rightarrow He^5 + n \rightarrow He^4 + 2n$ (in this rough approximation, we consider only nucleons). The typical energy value will be about $10\ Mev\approx 10^{-5}\ erg$, and the energy-time uncertainty relation will give $dt\approx 10^{-22}\ sec$. The whole process takes about $1.6 \ sec$, from where $N=t/dt\approx 10^{22}\approx 2^{73}$, and if quantum mechanics can be continued to nuclear processes such as the decay of Helium-6 isotope to the stable isotope $He^4$, GSA must work well already at $73$ qubits.

The decay of $He^6$ from the viewpoint of quantum mechanics is a very complex process. It is possible to consider only its last stage when one neutron is split off from the stable nucleus of $He^4$. It takes approximately $10^{-11}\ sec$. For it, estimates similar to the above will give approximately $36$ qubits of a reliable implementation of GSA, which is already less realistic, but the corresponding value of $Q\approx 2^{36}$ can yet be verified on GSA experiments. 

   The adoption of the accuracy-complexity uncertainty hypothesis thus directly links the question of the applicability of quantum theory to real microprocesses and the implementation of GSA. The implementation of GSA thus becomes a central issue of quantum theory and the theory of complex systems as such.

\section{Conclusion}

We have described the structure of the operating system of a quantum computer needed to perform its main task - to control the dynamics at the quantum level. Quantum device drivers here are based on the Tavis-Cummings-Hubbard model and its modifications involving charge movements. The central part of a quantum operating system should be a classical algorithm that simulates the quantum dynamics of many charges and fields.

The operating system of a quantum computer, being a classical program executed on a supercomputer, is subject to the uncertainty relation ''the accuracy of the description of the wave function - the complexity of the quantum system'', the constant of which can be found by analyzing the behavior of the state of a quantum computer implementing Grover algorithm. The number of qubits for which this algorithm will operate normally directly determines this constant.

 In the standard formulation of quantum theory, this constant is infinity - the maximum number of possible states in an entangled quantum superposition that cannot be simplified through canonical transformations. This makes it impossible to directly find the wave function of a system of many particles using statistical analysis. Thus, in the standard formulation, the quantum theory of many bodies can only be verified by implementing Grover fast quantum algorithm on thousands of qubits, which seems doubtful in light of the current state of quantum computer experiments.

 Time and energy of specific processes allow to estimate approximately the value of this constant. For example, the implementation of the Grover algorithm for about 70 qubits is necessary in order for nuclear processes to be described by the methods of quantum theory in the same way as the movement of an electron in an atom. The possibility of implementing Grover algorithm for about 12 qubits follows from the possibility of a very accurate quantum description of the dynamics of an electron in an atom. This determines the importance of experiments on the implementation of this algorithm.

This uncertainty relation actually gives a bound on the applicability of quantum theory in terms of algorithmic complexity, and this bound is quite achievable in experiments.

\section{Acknowledgements}

The work is supported by The Russian Foundation for Basic Research, grant a-18-01-00695.

\end{document}